\documentstyle[12pt]{article}
\begin{document}
\def\hybrid{\topmargin 0pt      \oddsidemargin 0pt
        \headheight 0pt \headsep 0pt
        \textheight 9in         
        \textwidth 6.25in       
        \marginparwidth .875in
        \parskip 5pt plus 1pt   \jot = 1.5ex}
\catcode`\@=11
\def\marginnote#1{}
\hyphenation{bo-so-ni-zed}
\newcount\hour
\newcount\minute
\newtoks\amorpm
\hour=\time\divide\hour by60 \minute=\time{\multiply\hour by60 \global\advance\minute
by-\hour}\edef\standardtime{{\ifnum\hour<12
\global\amorpm={am}%
        \else\global\amorpm={pm}\advance\hour by-12 \fi
        \ifnum\hour=0 \hour=12 \fi
        \number\hour:\ifnum\minute<10
0\fi\number\minute\the\amorpm}} \edef\militarytime{\number\hour:\ifnum\minute<10
0\fi\number\minute}
\def\draftlabel#1{{\@bsphack\if@filesw {\let\thepage\relax
   \xdef\@gtempa{\write\@auxout{\string
      \newlabel{#1}{{\@currentlabel}{\thepage}}}}}\@gtempa
   \if@nobreak \ifvmode\nobreak\fi\fi\fi\@esphack}
        \gdef\@eqnlabel{#1}}
\def\@eqnlabel{}
\def\@vacuum{}
\def\draftmarginnote#1{\marginpar{\raggedright\scriptsize\tt#1}}
\def\draft{\oddsidemargin -.5truein
        \def\@oddfoot{\sl preliminary draft \hfil
        \rm\thepage\hfil\sl\today\quad\militarytime}
        \let\@evenfoot\@oddfoot \overfullrule 3pt
        \let\label=\draftlabel
        \let\marginnote=\draftmarginnote

\def\@eqnnum{(\theequation)\rlap{\kern\marginparsep\tt\@eqnlabel}%
\global\let\@eqnlabel\@vacuum}  }


\def\numberbysection{\@addtoreset{equation}{section}
        \def\theequation{\thesection.\arabic{equation}}}

\def\underline#1{\relax\ifmmode\@@underline#1\else
 $\@@underline{\hbox{#1}}$\relax\fi}

\catcode`@=12 \relax

\numberbysection

\topmargin 0pt \advance \topmargin by -\headheight \advance \topmargin by -\headsep

\textheight 8.9in

\oddsidemargin 0pt \evensidemargin \oddsidemargin \marginparwidth 0.5in

\textwidth 6.5in

\topmargin -.6in

\def\rh{{\hat \rho}}
\def\alie{{\hat{\cal G}}}
\def\sect#1{\section{#1}}

\def\sd{D\!\!\!/}
\def\sp{\partial\!\!\!/}
\def\sa{A\!\!\!/}
\def\sb{b\!\!\!/}
\def\ss{s\!\!\!/}
\def\rf#1{(\ref{#1})}
\def\lab#1{\label{#1}}
\def\nonu{\nonumber}
\def\br{\begin{eqnarray}}
\def\er{\end{eqnarray}}
\def\be{\begin{equation}}
\def\ee{\end{equation}}
\def\eq{\!\!\!\! &=& \!\!\!\! }
\def\foot#1{\footnotemark\footnotetext{#1}}
\def\lb{\lbrack}
\def\rb{\rbrack}
\def\llangle{\left\langle}
\def\rrangle{\right\rangle}
\def\blangle{\Bigl\langle}
\def\brangle{\Bigr\rangle}
\def\llbrack{\left\lbrack}
\def\rrbrack{\right\rbrack}
\def\lcurl{\left\{}
\def\rcurl{\right\}}
\def\({\left(}
\def\){\right)}
\newcommand{\nit}{\noindent}
\newcommand{\ct}[1]{\cite{#1}}
\newcommand{\bi}[1]{\bibitem{#1}}
\def\lskip{\vskip\baselineskip\vskip-\parskip\noindent}
\relax

\def\tr{\mathop{\rm tr}}
\def\Tr{\mathop{\rm Tr}}
\def\v{\vert}
\def\bv{\bigm\vert}
\def\Bgv{\;\Bigg\vert}
\def\bgv{\bigg\vert}
\newcommand\partder[2]{{{\partial {#1}}\over{\partial {#2}}}}
\newcommand\funcder[2]{{{\delta {#1}}\over{\delta {#2}}}}
\newcommand\Bil[2]{\Bigl\langle {#1} \Bigg\vert {#2} \Bigr\rangle}  
\newcommand\bil[2]{\left\langle {#1} \bigg\vert {#2} \right\rangle} 
\newcommand\me[2]{\left\langle {#1}\bv {#2} \right\rangle} 
\newcommand\sbr[2]{\left\lbrack\,{#1}\, ,\,{#2}\,\right\rbrack}
\newcommand\pbr[2]{\{\,{#1}\, ,\,{#2}\,\}}
\newcommand\pbbr[2]{\lcurl\,{#1}\, ,\,{#2}\,\rcurl}
%
\def\a{\alpha}
\def\at{{\tilde A}^R}
\def\atc{{\tilde {\cal A}}^R}
\def\atcm#1{{\tilde {\cal A}}^{(R,#1)}}
\def\b{\beta}
\def\btil{{\tilde b}}
\def\dc{{\cal D}}
\def\d{\delta}
\def\D{\Delta}
\def\eps{\epsilon}
\def\bareps{{\bar \epsilon}}
\def\vareps{\varepsilon}
\def\fptil{{\tilde F}^{+}}
\def\fmtil{{\tilde F}^{-}}
\def\gh{{\hat g}}
\def\g{\gamma}
\def\G{\Gamma}
\def\grad{\nabla}
\def\h{{1\over 2}}
\def\l{\lambda}
\def\L{\Lambda}
\def\m{\mu}
\def\n{\nu}
\def\o{\over}
\def\om{\omega}
\def\O{\Omega}
\def\p{\phi}
\def\P{\Phi}
\def\pa{\partial}
\def\pr{\prime}
\def\pt{{\tilde \Phi}}
\def\qs{Q_{\bf s}}
\def\ra{\rightarrow}
\def\s{\sigma}
\def\S{\Sigma}
\def\t{\tau}
\def\th{\theta}
\def\Th{\Theta}
\def\tpp{\Theta_{+}}
\def\tmm{\Theta_{-}}
\def\tpg{\Theta_{+}^{>}}
\def\tms{\Theta_{-}^{<}}
\def\tp0{\Theta_{+}^{(0)}}
\def\tm0{\Theta_{-}^{(0)}}
\def\ti{\tilde}
\def\wti{\widetilde}
\def\jc{J^C}
\def\bj{{\bar J}}
\def\sj{{\jmath}}
\def\bsj{{\bar \jmath}}
\def\bp{{\bar \p}}
\def\vp{\varphi}
\def\vt{{\tilde \varphi}}
\def\faa{Fa\'a di Bruno~}
\def\ca{{\cal A}}
\def\cb{{\cal B}}
\def\ce{{\cal E}}
\def\cg{{\cal G}}
\def\cgh{{\hat {\cal G}}}
\def\ch{{\cal H}}
\def\chh{{\hat {\cal H}}}
\def\cl{{\cal L}}
\def\cm{{\cal M}}
\def\cn{{\cal N}}
\def\ns{N_{{\bf s}}}
\newcommand\sumi[1]{\sum_{#1}^{\infty}}   
\newcommand\fourmat[4]{\left(\begin{array}{cc}  
{#1} & {#2} \\ {#3} & {#4} \end{array} \right)}

%
\def\lie{{\cal G}}
\def\kmlie{{\hat{\cal G}}}
\def\dlie{{\cal G}^{\ast}}
\def\elie{{\widetilde \lie}}
\def\edlie{{\elie}^{\ast}}
\def\hlie{{\cal H}}
\def\flie{{\cal F}}
\def\wlie{{\widetilde \lie}}
\def\f#1#2#3 {f^{#1#2}_{#3}}
\def\winf{{\sf w_\infty}}
\def\win1{{\sf w_{1+\infty}}}
\def\hwinf{{\sf {\hat w}_{\infty}}}
\def\Winf{{\sf W_\infty}}
\def\Win1{{\sf W_{1+\infty}}}
\def\hWinf{{\sf {\hat W}_{\infty}}}
\def\Rm#1#2{r(\vec{#1},\vec{#2})}          
\def\OR#1{{\cal O}(R_{#1})}           
\def\ORti{{\cal O}({\widetilde R})}           
\def\AdR#1{Ad_{R_{#1}}}              
\def\dAdR#1{Ad_{R_{#1}^{\ast}}}      
\def\adR#1{ad_{R_{#1}^{\ast}}}       
\def\KP{${\rm \, KP\,}$}                 
\def\KPl{${\rm \,KP}_{\ell}\,$}         
\def\KPo{${\rm \,KP}_{\ell = 0}\,$}         
\def\mKPa{${\rm \,KP}_{\ell = 1}\,$}    
\def\mKPb{${\rm \,KP}_{\ell = 2}\,$}    
%
\def\rlx{\relax\leavevmode}
\def\inbar{\vrule height1.5ex width.4pt depth0pt}
\def\IZ{\rlx\hbox{\sf Z\kern-.4em Z}}
\def\IR{\rlx\hbox{\rm I\kern-.18em R}}
\def\IC{\rlx\hbox{\,$\inbar\kern-.3em{\rm C}$}}
\def\IN{\rlx\hbox{\rm I\kern-.18em N}}
\def\IO{\rlx\hbox{\,$\inbar\kern-.3em{\rm O}$}}
\def\IP{\rlx\hbox{\rm I\kern-.18em P}}
\def\IQ{\rlx\hbox{\,$\inbar\kern-.3em{\rm Q}$}}
\def\IF{\rlx\hbox{\rm I\kern-.18em F}}
\def\IG{\rlx\hbox{\,$\inbar\kern-.3em{\rm G}$}}
\def\IH{\rlx\hbox{\rm I\kern-.18em H}}
\def\II{\rlx\hbox{\rm I\kern-.18em I}}
\def\IK{\rlx\hbox{\rm I\kern-.18em K}}
\def\IL{\rlx\hbox{\rm I\kern-.18em L}}
\def\one{\hbox{{1}\kern-.25em\hbox{l}}}
\def\0#1{\relax\ifmmode\mathaccent"7017{#1}%
B        \else\accent23#1\relax\fi}
\def\omz{\0 \omega}
%
\def\ltimes{\mathrel{\vrule height1ex}\joinrel\mathrel\times}
\def\rtimes{\mathrel\times\joinrel\mathrel{\vrule height1ex}}
%
\def\mark{\noindent{\bf Remark.}\quad}
\def\prop{\noindent{\bf Proposition.}\quad}
\def\theor{\noindent{\bf Theorem.}\quad}
\def\name{\noindent{\bf Definition.}\quad}
\def\exam{\noindent{\bf Example.}\quad}
\def\proof{\noindent{\bf Proof.}\quad}
                %
                %

\def\BIO#1#2#3{{\sl Biometrika} {\bf #1} (#2) #3}
\def\TEC#1#2#3{{\sl Techno\-me\-trics} {\bf #1} (#2) #3}
\def\FRE#1#2#3{{\sl Fresenius J Anal Chem} {\bf #1} (#2) #3}
\def\ISR#1#2#3{{\sl International Statistical Review} {\bf #1} (#2) #3}
\def\JASA#1#2#3{{\sl Journal of the American Statistical Association} {\bf #1} (#2) #3}
                %
                \def\a{\alpha}
                \def\b{\beta}
                \def\ca{{\cal A}}
                \def\cm{{\cal M}}
                \def\cn{{\cal N}}
                \def\cf{{\cal F}}
                \def\d{\delta}
                \def\D{\Delta}
                \def\eps{\epsilon}
                \def\g{\gamma}
                \def\G{\Gamma}
                \def\vp{\varphi}
                \def\grad{\nabla}
                \def\h{ {1\over 2}  }
                \def\hc{\hat{c}}
                \def\hd{\hat{d}}
                \def\hg{\hat{g}}
                \def\/{\frac}
                \def\hp{ {+{1\over 2}}  }
                \def\hm{ {-{1\over 2}}  }
                \def\k{\kappa}
                \def\l{\lambda}
                \def\L{\Lambda}
                \def\lg{\langle}
                \def\m{\mu}
                \def\n{\nu}
                \def\o{\omega}
                \def\O{\Omega}
                \def\p{\phi}
                \def\pa{\partial}
                \def\pr{\prime}
                \def\qq{\qquad}
                \def\ra{\rightarrow}
                \def\rh{\rho}
                \def\vp{\varphi}
                \def\rg{\rangle}
                \def\s{\sigma}
                \def\t{\tau}
                \def\th{\theta}
                \def\ti{\tilde}
                \def\u{\upsilon}
                \def\wti{\widetilde}
                \def\inte{\int dx }
                \def\xb{\bar{x}}
                \def\yb{\bar{y}}
                \def\({\Big(}
                \def\){\Big)}
                \def\[{\Big[}
                \def\]{\Big]}

                \def\rlx{\relax\leavevmode}
                \def\inbar{\vrule height1.5ex width.4pt depth0pt}
                \def\IZ{\rlx\hbox{\sf Z\kern-.4em Z}}
                \def\IR{\rlx\hbox{\rm I\kern-.18em R}}
                \def\IC{\rlx\hbox{\,$\inbar\kern-.3em{\rm C}$}}
                \def\IN{\rlx\hbox{\rm I\kern-.18em N}}
                \def\IO{\rlx\hbox{\,$\inbar\kern-.3em{\rm O}$}}
                \def\IP{\rlx\hbox{\rm I\kern-.18em P}}
                \def\IQ{\rlx\hbox{\,$\inbar\kern-.3em{\rm Q}$}}
                \def\IF{\rlx\hbox{\rm I\kern-.18em F}}
                \def\IG{\rlx\hbox{\,$\inbar\kern-.3em{\rm G}$}}
                \def\IH{\rlx\hbox{\rm I\kern-.18em H}}
                \def\II{\rlx\hbox{\rm I\kern-.18em I}}
                \def\IK{\rlx\hbox{\rm I\kern-.18em K}}
                \def\IL{\rlx\hbox{\rm I\kern-.18em L}}
                \def\one{\hbox{{1}\kern-.25em\hbox{l}}}
                \def\0#1{\relax\ifmmode\mathaccent"7017{#1}%
                B        \else\accent23#1\relax\fi}
                \def\omz{\0 \omega}
                %

                \def\tr{\mathop{\rm tr}}
                \def\Tr{\mathop{\rm Tr}}
                \def\partder#1#2{{\partial #1\over\partial #2}}
                \def\ds{{\cal D}_s}
                \def\wtwo{{\wti W}_2}
\vspace{.2in}
\begin{center}
{\large\bf Homoscedastic controlled calibration model}
\end{center}

\vspace{0.5in}

\begin{center}

Betsab\'e G. Blas Achic$^{a}$, M\^onica C. Sandoval  $^{b}$ and Olga Satomi Yoshida$^{c}$

\vspace{.5 cm} \small

\par \vskip .1in \noindent
$^{a,}$$^{b}$Departamento de Estat\'{\i}stica, Universidade de S\~ao Paulo,
S\~ao Paulo, Brasil\\
$^{c}$ Centro de Metrologia de Fluidos, Instituto de Pesquisas Tecnol\'ogicas, S\~ao Paulo, Brasil\\
\normalsize
\end{center}
\vspace{1 cm}

\begin{abstract}
In the context of the usual calibration model, we consider the case in which the
independent variable is unobservable, but a pre-fixed value on its surrogate is
available. Thus, considering controlled variables and assuming that the measurement errors have equal variances we propose a new calibration model. Likelihood based methodology is used to estimate the
model parameters and the Fisher information matrix is used to construct a
confidence interval for the unknown value of the regressor variable. A simulation
study is carried out to asses the effect of the measurement error on the estimation
of the parameter of interest. This new approach is illustrated with an example.
\end{abstract}




\par \vskip .3in \noindent
\vspace{1 cm}

\section{Introduction}
In the first stage of a calibration problem, a pair of data sample $(x_{i},Y_{i}),\,i=1,2,\cdots n$ is observed. In the second stage, it is
observed one or more values, which are the responses corresponding to a
single unknown value of the regressor variable, $X_{0}$.
The first and second stage equations of the usual linear calibration model are defined, respectively, as
\begin{eqnarray}
\label{m1}
Y_{i}&=&\alpha+\beta x_{i}+\epsilon_{i},\,\,\,\,\,i=1,2\cdots,n,\\
\label{m2}
Y_{0i}&=&\alpha+\beta X_{0}+\epsilon_{i},\,\,\,\,\,i=n+1,n+2,\cdots,n+k.
\end{eqnarray}
It is considered the following assumptions:
\begin{itemize}
  \item $x_{1},x_{2},\cdots,x_{n}$ take fixed values, which are considered as true values.
  \item $\epsilon_{1},\epsilon_{2},\cdots,\epsilon_{n+k}$ are independent and normally distributed with mean 0 and variance $\sigma_{\epsilon}^{2}$.
\end{itemize}
The model parameters are $\alpha,\beta,X_{0}$ and $\sigma_{\epsilon}^{2}$ and the
main interest is to estimate the quantity $X_{0}$.

The maximun likelihood estimators of the usual calibration model are given by
\begin{eqnarray}
\label{xo}
\hat{\alpha}&=&\bar{Y}-\hat{\beta}\bar{x},\,\,\,\,\,\,\,\,\,\,\,\,\,\,\,\,\hat{\beta}=\frac{S_{xY}}{S_{xx}},\,\,\,\,\,\,\,\,\,\,\,\,\,\,\,\hat{X}_{0}=\frac{\bar{Y}_{0}-\hat{\alpha}}{\hat{\beta}},\\
\label{se}
\sigma_{\epsilon}^{2}&=&\frac{1}{n+k}\big[\sum_{i=1}^{n}(Y_{i}-\hat{\alpha}-\hat{\beta}x_{i})^{2}+\sum_{i=n+1}^{n+k}(Y_{0i}-\bar{Y}_{0})^{2}\big],
\end{eqnarray}
where
\begin{eqnarray}
\nonumber \bar{x}&=&\frac{1}{n}\sum_{i=1}^{n}x_{i},\,\,\,\,
\bar{Y}=\frac{1}{n}\sum_{i=1}^{n}Y_{i},\,\,\,\,
S_{xY}=\frac{1}{n}\sum_{i=1}^{n}(x_{i}-\bar{x})(Y_{i}-\bar{Y}),\\\nonumber
S_{xx}&=&\frac{1}{n}\sum_{i=1}^{n}(x_{i}-\bar{x})^{2},\,\,\,\,
\bar{Y}_{0}=\frac{1}{n}\sum_{i=n+1}^{n+k}Y_{0i}.
\end{eqnarray}
In \cite{eura} an approximate expression is derived for the variance of the estimator $\hat{X}_{0}$, which is derived through the propagation error law. Another approximation for the variance of $\hat{X}_{0}$ is given by the Fisher information of $\theta=(\alpha,\beta,X_{0},\sigma_{\epsilon}^{2})$ which, after some length algebraic manipulations, it can be shown to be given by
 \br I(\theta)= \frac{1}{\sigma_{\epsilon}^{2}}
\left(\begin{array}{clrrr}
     n+k &  kX_{0}+n\bar{x}           &k\beta  &0 \\
     kX_{0}+n\bar{x}            &   kX^{2}_{0}+\sum_{i=1}^{n}x_{i}^{2}  & \k\beta X_{0}& 0\\
      k\beta &  \k\beta X_{0} & k\beta^{2} & 0\\
      0 & 0 & 0 & \frac{n+k}{2\sigma_{\epsilon}^{2}}
\end{array}\right).
\er

The maximum likelihood estimator of
$\hat{\theta}=(\hat{\alpha},\hat{\beta},\hat{X_{0}},\hat{\sigma_{\epsilon}^{2}})$ has
approximately normal distribution with mean $\theta$ and covariance matrix
$I(\theta)^{-1}$, when $k=qn,\,q\in Q^{+}$ and $n\longrightarrow\infty$. Thus, the approximation of order $n^{-1}$ for the variance of $\hat{X}_{0}$ is given by

\begin{equation}
 \label{veurachem}
V_{1}(\hat{X}_{0})=\frac{\sigma_{\epsilon}^{2}}{\beta^{2}}\left[\frac{1}{k}+\frac{1}{n}+\frac{(\bar{X}-X_{0})^{2}}{nS_{xx}}\right].
\end{equation}

On the other hand, in \cite{shukla} the size $k$ of the second stage is  considered
fixed, so that expanding $\hat{X}_{0}$ in Taylor series around the point ($\alpha,\beta$)
and ignoring terms of order less than $n^{-2}$, we can find the following
approximations for the bias and variance of $\hat{X}_{0}$, respectively,
\begin{eqnarray}
 \label{vicioshuk}
Bias(\hat{X}_{0})&=&\frac{\sigma_{\epsilon}^{2}(X_{0}-\bar{x})}{n\beta^{2}S_{xx}},\\
\label{vshuk}
V_{2}(\hat{X}_{0})&=&\frac{\sigma_{\epsilon}^{2}}{\beta^{2}}\left[\frac{1}{k}+\frac{1}{n}+\frac{(\bar{X}-X_{0})^{2}}{nS_{xx}}+\frac{3\sigma_{\epsilon}^{2}}{nk\beta^{2}S_{xx}}\right].
\end{eqnarray}

In order to construct a confidence interval for $X_{0}$, we consider that
\begin{equation}
\label{pibotep}
\frac{\hat{X}_{0}-X_{0}}{\sqrt{\hat{V}(\hat{X}_{0})}}\stackrel{D}{\longrightarrow
}N(0,1),
\end{equation}
where $\hat{V}(\hat{X}_{0})$ is the estimated variance computed
according to (\ref{veurachem}) or (\ref{vshuk}). Hence, the approximated
confidence interval for $X_{0}$ with a confidence level $(1-\alpha)$, is given by
\begin{equation}
\label{uinterv}
\left(\hat{X}_{0}-z_{\frac{\alpha}{2}}\sqrt{\hat{V}(\hat{X}_{0})},\hat{X}_{0}+z_{\frac{\alpha}{2}}\sqrt{\hat{V}(\hat{X}_{0})}\right),
\end{equation}
where $z_{\frac{\alpha}{2}}$ is the quantile of order
$(1-\frac{\alpha}{2})$ of the standard normal distribution.\\

The usual calibration problem has been discussed in the literature for several
decades (see \cite{berk69}-\cite{tell}). An illustration of this
model is presented for example in \cite{dunsmore}. We can find a review of the
literature on statistical calibration in \cite{osbourne}, where some approaches to
the solution of the calibration problem are summarized.

This model encounters applications in different areas, but it is not well
suited in some instances as, for example, in chemical analysis, where the preparation
process of standard solutions are subject to measurement error (\cite{eura}).

There exists some situations, as mentioned above, where the independent variable,
$x_{i}$, is measured with error. In this case, \cite{berk50} defines two types of
observations: {\sl controlled} and {\sl uncontrolled}.

In the {\sl  uncontrolled} situation, the usual procedure to obtain the true value of the independent variable $x_{i}$ generates an error and the observed value is
 \br \label{e1} X_{i}=x_{i}+\delta_{i}, \,\,\,\,\,\,i=1,\cdots,n. \er

We have that $x_{i}$ is an unknown quantity, $\delta_{i}$ is a measurement
error and  $X_{i}$ is a random variable. Assuming that $x_{i}$ is a parameter the model defined by (\ref{m1}) and (\ref{e1}) is named as functional model (\cite{fuller}). In this case there exists correlation between the model error and the variable $X_{i}$. Assuming that $x_{i}$ is a random variable the model (\ref{m1}) and (\ref{e1}) is called as structural model (\cite{fuller}). On the other hand, the model defined by (\ref{m1}), (\ref{m2}) and (\ref{e1}) is called as the functional or structural calibration model if $x_{i}$ is assumed as a parameter or a random variable, respectively (\cite{lima}).

The {\sl controlled} observation is defined by a pre-fixed value $X_{i}$ according to the
experimenter
convenience and a procedure is established in order to attain the pre-fixed value.
The experiment gives the unobserved $x_{i}$ and it is such that
 \be  \label{e2} x_{i}= X_{i}-\delta _{i}, \,\,\,\,\,\,i=1,\cdots,n. \ee
In this case, the fixed quantity is $X_{i}$, the measurement error is $\delta_{i}$
and $x_{i}$ is the random variable. The model (\ref{m1}) and (\ref{e2}) is known as
Berkson regression model (\cite{cheng}). Notice that the model error and the
quantity $X_{i}$ are independent. The model  defined by (\ref{m1}), (\ref{m2}) and
(\ref{e2}) has not been considered before in the measurement error literature and in
this work it will be called as the {\sl controlled calibration model}.

In the calibration model defined by (\ref{m1}), (\ref{m2}) and  (\ref{e1}), the values of the regressor, $X_{i}$, from the first stage are randomly generated, whereas in the controlled calibration model, (\ref{m1}), (\ref{m2}) and  (\ref{e2}), they are assumed as pre-fixed by the experimenter.

This work is organized as follows. In Section 2, we derive the maximum likelihood
estimators of the homoscedastic controlled calibration model by considering both
cases: $\sigma_{\delta}^{2}$ {\sl unknown} and {\sl known}. In Section 3, a
simulation study is undertaken to investigate the sensitivity of parameter estimates
of the proposed model. In Section 4, an example is presented to illustrate our new
approach. In Section 5, the concluding remark is presented.

\section{Parameter estimation}
In this section we study the controlled calibration model. From  the equations (\ref{m1}), (\ref{m2}) and  (\ref{e2}) we can write
\br \label{mc1}
Y_{i}&=&\alpha+\beta X_{i}+(\epsilon_{i}-\beta\delta_{i}),\,\,\,\,\,i=1,2\cdots,n,\\
\label{mc2}
Y_{0i}&=&\alpha+\beta X_{0}+\epsilon_{i},\,\,\,\,\,i=n+1,n+2,\cdots,n+k.
\er
with the following assumptions for the random errors
\begin{itemize}
  \item $\epsilon_{i}$ are independent $N(0,\sigma_{\epsilon}^{2})$ random variables. 
  \item E($\delta_{i}$)=0, V($\delta_{i}$)= $\sigma_{\delta_{i}}^{2}$.
  \item cov($\delta_{i},\delta_{j}$)=0 for any $i\neq j$.
  \item cov($\epsilon_{i},\delta_{j}$)=0 for all $i,j$.
\end{itemize}

Some comments are in order here. The variable $X_{i}$ in (\ref{mc1}) is
controlled and the error model $(\epsilon_{i}-\beta\delta_{i})$ is independent of
$X_{i}$. The error model in (\ref{mc2}) is only in function of error measure $\epsilon_{i}$ related to $Y_{0i}$, this model assume that there is not error in the preparation sample related to parameter $X_{0}$. We define the homoscedastic controlled calibration model by considering that
the errors $\delta_{i}$ are independent and normally distributed with mean 0 and constant
variance, $\sigma_{\delta}^{2}$. The study of this model is carried out
following similar analysis to the usual calibration model as summarized above.

The maximum likelihood estimator for the homoscedastic controlled
calibration model is derived in the following. The logarithm of the
likelihood function is given by:
\begin{eqnarray}
\label{L} \nonumber
l(\alpha,\beta,X_{0},\sigma_{\epsilon}^{2},\sigma_{\delta}^{2})&\propto&-\frac{n}{2}log(\sigma_{\epsilon}^{2}+
\beta^{2}\sigma_{\delta}^{2})-
\frac{k}{2}log(\sigma_{\epsilon}^{2})-\\\nonumber
&&\frac{1}{2}\bigg[\frac{1}{\sigma_{\epsilon}^{2} +
\beta^{2}\sigma_{\delta}^{2}}\sum_{i=1}^{n}(Y_{i}-\alpha-\beta
X_{i})^{2}+\\
&&\frac{1}{\sigma_{\epsilon}^{2}}\sum_{i=n+1}^{n+k}(Y_{0i}-\alpha-\beta
X_{0})^{2}\bigg].
\end{eqnarray}
Solving $\partial l/\partial\alpha=0$ and $\partial
l/\partial X_{0}=0$ we have the maximum likelihood estimator of
$\alpha$ and $X_{0}$, which are given, respectively, by
\br
\hat{\alpha}=\bar{Y}-\hat{\beta}\bar{X}\,\,\,\,\,\,\mbox{and}\,\,\,\,\,\,\hat{X}_{0}=\frac{\bar{Y}_{0}-\hat{\alpha}}{\hat{\beta}}.\label{solu}\er
From (\ref{L}) and (\ref{solu}), it follows that the likelihood for $(\beta,\sigma^{2}_{\epsilon},\sigma^{2}_{\delta})$ can be written as
\begin{eqnarray}
\label{Ls}\nonumber
l(\beta,\sigma_{\epsilon}^{2},\sigma_{\delta}^{2})\propto&-&\frac{n}{2}log(\sigma_{\epsilon}^{2}+
\beta^{2}\sigma_{\delta}^{2})-\frac{k}{2}log(\sigma_{\epsilon}^{2})-\frac{1}{2}\bigg[\frac{1}{\sigma_{\epsilon}^{2}
+ \beta^{2}\sigma_{\delta}^{2}}\sum_{i=1}^{n}[(Y_{i}-\bar{Y})\\
&-&\beta(X_{i}-\bar{X})]^{2}+\frac{1}{\sigma_{\epsilon}^{2}}\sum_{i=n+1}^{n+k}(Y_{0i}-\bar{Y_{0}})^{2}\bigg].
\end{eqnarray}

Next, we consider  two cases for $\sigma_{\delta}^{2}$. Firstly, we obtain the maximum likelihood estimator of $\beta,\,\,\sigma^{2}_{\epsilon}$ and
$\sigma^{2}_{\delta}$ from (\ref{Ls}). In the second case we assume that the variance
$\sigma_{\delta}^{2}$ is known and obtain the maximum likelihood
estimators for $\beta$ and $\sigma^{2}_{\epsilon}$.\\
\\
{\bf\large Case 1: {\sl unknown} variance $\sigma_{\delta}^{2}$}\\

Taking the partial derivative of (\ref{Ls}) with respect to $\beta,\,\,\sigma^{2}_{\epsilon}$ and $\sigma_{\delta}^{2}$ and
equating to zero we obtain, respectively,
\begin{eqnarray}
\label{eb}
\hat{\beta}\hat{\sigma}_{\delta}^{2}(\hat{\sigma}_{\epsilon}^{2}+\hat{\beta}^{2}\hat{\sigma}_{\delta}^{2}-S_{YY}+\hat{\beta}S_{XY})&=&(S_{XY}-\hat{\beta} S_{XX})\hat{\sigma}_{\epsilon}^{2},\\
\label{ed}
\hat{\sigma}_{\epsilon}^{2}+\hat{\beta}^{2}\hat{\sigma}_{\delta}^{2}&=&S_{YY}-2\hat{\beta} S_{XY}+\hat{\beta}^{2}S_{XX},\\
\nonumber
\frac{kS_{Y_{0}Y_{0}}}{(\hat{\sigma}_{\epsilon}^{2})^{2}}-\frac{k}{\hat{\sigma}_{\epsilon}^{2}}&=&\frac{n}{\hat{\sigma}_{\epsilon}^{2}+\hat{\beta}^{2}\hat{\sigma}_{\delta}^{2}}-\frac{n(S_{YY}-2\hat{\beta}
S_{XY}+\hat{\beta}^{2}S_{XX})}{(\hat{\sigma}_{\epsilon}^{2}+\hat{\beta}^{2}\hat{\sigma}_{\delta}^{2})^{2}},\\\label{ee}
\end{eqnarray}
where
$S_{XX}=\frac{1}{n}\sum^{n}_{i=1}(X_{i}-\bar{X})^{2},S_{XY}=\frac{1}{n}\sum^{n}_{i=1}(X_{i}-\bar{X})(Y_{i}-\bar{Y}),S_{YY}=\frac{1}{n}\sum^{n}_{i=1}(Y_{i}-\bar{Y})^{2}$ and  $S_{Y_{0}Y_{0}}=\frac{1}{k}\sum^{n+k}_{i=n+1}(Y_{0i}-\bar{Y_{0}})^{2}$, and the relevant estimator notation has been introduced. From (\ref{eb}) and (\ref{ed}) we have the following equations:
$$(\hat{\beta} S_{XX}-S_{XY})(S_{YY}-2\hat{\beta} S_{XY}+\hat{\beta}^{2}S_{XX})=0,$$
hence
\begin{eqnarray}
\label{err} \hat{\beta} S_{XX}-S_{XY}&=&0\,\,\,\,\,\,\,\,\mbox{or}\\
\label{eri} S_{YY}-2\hat{\beta} S_{XY}+\hat{\beta}^{2}S_{XX}&=&0.
\end{eqnarray}
Therefore, from (\ref{err}), we have that $\hat{\beta} =S_{XY}/S_{XX}$.
But, according to the Cauchy-Schwarz inequality, $S_{XX}S_{YY}\geq S_{XY}^{2}$, hence  (\ref{eri}) has real roots if and only if $Y_{i}=cX_{i}$, where $c$ is a
constant.


The estimator of $\sigma_{\delta}^{2}$ can be obtained from the equation
(\ref{ed})
$$\hat{\sigma}_{\delta}^{2}=\frac{(S_{YY}-2\hat{\beta} S_{XY}+\hat{\beta}^{2}S_{XX})-\hat{\sigma}_{\epsilon}^{2}}{\hat{\beta}^{2}}.$$

Likewise, from equations (\ref{ed}) and (\ref{ee}) we obtain the estimator
of the variance $\sigma_{\epsilon}^{2}$
\begin{equation}
\label{sigmades} \hat{\sigma}_{\epsilon}^{2}=S_{Y_{0}Y_{0}}.
\end{equation}

In order to find the variance of $\hat{X}_{0}$, we need to derive the Fisher information matrix of
$\theta=(\alpha,\beta,X_{0},\sigma_{\delta}^{2},\sigma_{\epsilon}^{2})$, which can be shown to be given by
\br I(\theta)= \left(\begin{array}{ccccc}
     \frac{n}{\gamma}+\frac{k}{\sigma_{\epsilon}^{2}} & \frac{n\bar{X}}{\gamma}+ \frac{kX_{0}}{\sigma_{\epsilon}^{2}}            &\frac{k\beta}{\sigma_{\epsilon}^{2}} &0 &0 \\
     \frac{n\bar{X}}{\gamma}+ \frac{kX_{0}}{\sigma_{\epsilon}^{2}} &        \frac{\sum_{i=1}^{n}X_{i}^{2}}{\gamma}+\frac{2n\beta^{2}\sigma_{\delta}^{4}}{\gamma^{2}}+\frac{kX_{0}^{2}}{\sigma_{\epsilon}^{2}}  & \frac{k\beta X_{0}}{\sigma_{\epsilon}^{2}}  & \frac{n\beta^{3}\sigma_{\delta}^{2}}{\gamma^{2}} & \frac{n\beta\sigma_{\delta}^{2}}{\gamma^{2}}\\
      \frac{k\beta}{\sigma_{\epsilon}^{2}} &  \frac{k\beta X_{0}}{\sigma_{\epsilon}^{2}} &\frac{k\beta^{2}}{\sigma_{\epsilon}^{2}} &0 &0 \\
      0 & \frac{n\beta^{3}\sigma_{\delta}^{2}}{\gamma^{2}} & 0& \frac{n\beta^{4}}{2\gamma^{2}} &\frac{n\beta^{2}}{2\gamma^{2}}\\
      0 & \frac{n\beta\sigma_{\delta}^{2}}{\gamma^{2}} & 0& \frac{n\beta^{2}}{2\gamma^{2}} &\frac{n}{2\gamma^{2}}+\frac{k}{2\sigma_{\epsilon}^{4}}\\
\end{array}\right),\nonumber
\er
where
\begin{equation}
\label{gama} \gamma=\beta^{2}\sigma_{\delta}^{2}+\sigma_{\epsilon}^{2}.
\end{equation}
When $k=qn$, $q\in Q^{+}$ and $n\longrightarrow\infty$, the estimator
$\hat\theta$ is approximately normally distributed with mean $\theta$ and variance $I(\theta)^{-1}$, thus we have that the
approximate variance to order $n^{-1}$ for $\hat{X}_{0}$ is given by
\begin{equation}
\label{vhomo}
V_{1}(\hat{X}_{0})=\frac{\sigma_{\epsilon}^{2}}{\beta^{2}}\left[\frac{1}{k}+\frac{\gamma}{n\sigma_{\epsilon}^{2}}+\frac{\gamma}{\sigma_{\epsilon}^{2}}\frac{(\bar{X}-X_{0})^{2}}{nS_{XX}}\right].
\end{equation}

Considering $k$ fixed and expanding $\hat{X}_{0}$
in a Taylor series around $(\alpha,\beta)$ and ignoring terms of order less than
$n^{-2}$, it can be shown that the bias and  variance of $\hat{X}_{0}$ (the proof is given in Appendix A), are given by
\begin{eqnarray}
\label{Biasxoet}
Bias(\hat{X}_{0})&=&\frac{\gamma(\bar{X}-X_{0})}{n\beta^{2}S_{XX}},\\
\label{vxoet}
V_{2}(\hat{X}_{0})&=&\frac{\sigma_{\epsilon}^{2}}{\beta^{2}}\left[\frac{1}{k}+\frac{\gamma}{n\sigma_{\epsilon}^{2}}+\frac{\gamma(\bar{X}-X_{0})^{2}}{n\sigma_{\epsilon}^{2}S_{XX}}+\frac{3\gamma}{nk\beta^{2}S_{XX}}\right].
\end{eqnarray}
We can observe that the estimator  of $X_{0}$ is
biased, but it is asymptotically unbiased.

With relation to the variance of the estimator $\hat{X}_{0}$, let us
notice that when $k=qn,\,\,q\in Q^{+}$, and ignoring the terms of
order less than $n^{-1}$ the variance in (\ref{vxoet})
coincide with the variance given in (\ref{vhomo}), which was found
through the Fisher information. Equation (\ref{vhomo}) consider large sample sizes in the first and second stage ($n$ and $k$),
whereas (\ref{vxoet}) consider large sample sizes in the
first stage and a fixed sample size in the second stage.

Notice that when $\sigma_{\delta}^{2}=0$, (\ref{vhomo}) and
(\ref{vxoet}) coincide with (\ref{veurachem}) and (\ref{vshuk}) of the usual model,
respectively.
\\

{\bf\large Caso 2: {\sl known} variance $\sigma_{\delta}^{2}$}\\

Assuming now that $\sigma_{\delta}^{2}$ is known and equating to
zero the partial derivative of (\ref{Ls}) with respect to the parameters $\beta$ and $\sigma_{\epsilon}^{2}$, we have the following
equations, respectively,
\begin{eqnarray}
\label{ebsd}
\hat{\beta}\sigma_{\delta}^{2}(\hat{\sigma}_{\epsilon}^{2}+\hat{\beta}^{2}\sigma_{\delta}^{2}-S_{YY}+\hat{\beta}S_{XY})&=&(S_{XY}-\hat{\beta} S_{XX})\hat{\sigma}_{\epsilon}^{2}\,\,\,\,\,\,\,\mbox{and}\,\,\,\,\,\\
\nonumber
\frac{kS_{Y_{0}Y_{0}}}{(\hat{\sigma}_{\epsilon}^{2})^{2}}-\frac{k}{\hat{\sigma}_{\epsilon}^{2}}&=&\frac{n}{\hat{\sigma}_{\epsilon}^{2}+\hat{\beta}^{2}\sigma_{\delta}^{2}}-\frac{S_{YY}-2\hat{\beta}
S_{XY}+\hat{\beta}^{2}S_{XX}}{(\hat{\sigma}_{\epsilon}^{2}+\hat{\beta}^{2}\sigma_{\delta}^{2})^{2}}.\\\label{eesd}
\end{eqnarray}
The estimates of $\beta$ and $\sigma_{\epsilon}^{2}$ are obtained
using some iterative method to solve (\ref{ebsd}) and
(\ref{eesd}).

Similarly, as in Case 1, the Fisher information matrix of
$\theta=(\alpha,\beta,X_{0},\sigma_{\epsilon}^{2})$ is given by
\br
I(\theta)= \left(\begin{array}{clrrr}
     \frac{n}{\gamma}+\frac{k}{\sigma_{\epsilon}^{2}} & \frac{n\bar{X}}{\gamma}+ \frac{kX_{0}}{\sigma_{\epsilon}^{2}}            &\frac{k\beta}{\sigma_{\epsilon}^{2}}  &0 \\
     \frac{n\bar{X}}{\gamma}+ \frac{kX_{0}}{\sigma_{\epsilon}^{2}} &        \frac{\sum_{i=1}^{n}X_{i}^{2}}{\gamma}+\frac{2n\beta^{2}\sigma_{\delta}^{4}}{\gamma^{2}}+\frac{kX_{0}^{2}}{\sigma_{\epsilon}^{2}}  & \frac{k\beta X_{0}}{\sigma_{\epsilon}^{2}}  & \frac{n\beta\sigma_{\delta}^{2}}{\gamma^{2}}\\
      \frac{k\beta}{\sigma_{\epsilon}^{2}} &  \frac{k\beta X_{0}}{\sigma_{\epsilon}^{2}} &\frac{k\beta^{2}}{\sigma_{\epsilon}^{2}} &0 \\
      0 & \frac{n\beta\sigma_{\delta}^{2}}{\gamma^{2}} & 0& \frac{n}{2\gamma^{2}}+\frac{k}{2\sigma_{\epsilon}^{4}}\\
\end{array}\right),\er
where $\gamma$ is defined in (\ref{gama}).

The large sample variance of $\hat{X}_{0}$ follows by inverting the
Fisher information matrix and is given by
\br
\label{vxoif}
V(\hat{X}_{0})&=&\frac{\sigma_{\epsilon}^{2}}{\beta^{2}}\left[\frac{1}{k}+\frac{\gamma}{n\sigma_{\epsilon}^{2}}+
\frac{\gamma}{\sigma_{\epsilon}^{2}}E\right],
\er
where,
$$E=\frac{nX_{0}^{2}\sigma_{\epsilon}^{4}+kX_{0}^{2}\gamma^{2}-2nX_{0}\bar{X}\sigma_{\epsilon}^{4}-2kX_{0}\bar{X}\gamma^{2}+n\bar{X}^{2}\sigma_{\epsilon}^{4}+k\bar{X}^{2}\gamma^{2}}{(n\sigma_{\epsilon}^{4}+k\gamma^{2})\sum_{i=1}^{n}X_{i}^{2}+2nk\beta^{2}\gamma\sigma_{\delta}^{4}-n^{2}\bar{X}^{2}\sigma_{\epsilon}^{4}-nk\bar{X}^{2}\gamma^{2}}.$$

Notice that if $\sigma_{\delta}^{2}=0$, the expression (\ref{vxoif}) is
reduced to (\ref{veurachem}).\\

To construct a  confidence interval for $X_{0}$, for both cases $\sigma_{\delta}^{2}$ {\sl unknown} and {\sl known}, we consider the interval (\ref{uinterv}), where $\hat{V}(\hat{X}_{0C})$ is the estimated variance that follows from (\ref{vhomo}), (\ref{vxoet}) or (\ref{vxoif}).

\section{Simulation study}
In this section we present a simulation study for both cases of the homoscedastic
controlled calibration model: $\sigma_{\delta}^{2}$ known and unknown. The objetive
of this section is to study the performance of the estimators of the proposed
model (Proposed-M) and verify the impact by considering erratically the usual model
(Usual-M).

It was considered 5000 samples generated from the homoscedastic controlled
calibration model. In all samples, the value of the parameters $\alpha$ and $\beta$
were 0.1 and 2, respectively. The range of values for the controlled variable was
[0,2]. The fixed values for the controlled variable were
$x_{1}=0,\,x_{i}=x_{i-1}+\frac{2}{n-1},\,i=2,\cdots,n,$ and the parameter values
$X_{0}$ were 0.01 (extreme inferior value), 0.8 (near to the central value) and 1.9
(extreme superior value). It was considered $\sigma_{\epsilon}^{2}=0.04$ and the
parameter values of $\sigma_{\delta}^{2}$ were 0.01 and 0.1, which are named,
respectively, as small and large variances. For the first and second stages we consider the sample of sizes $n=5,\,20,\,100$ and $k=2,\,20,\,100$, respectively.

The empirical mean bias is given by
$\sum_{j=1}^{5000}(\hat{X}_{0}-X_{0})/5000$  and the empirical
mean squared error (MSE) is given by
$\sum_{j=1}^{5000}(\hat{X}_{0}-X_{0})^{2}/5000$. The mean estimated variance
 of $\hat{X}_{0}$ is given by
$\sum_{j=1}^{5000}\hat{V}(\hat{X}_{0})/5000$, with
$\hat{V}(\hat{X}_{0})=\hat{V}_{1}(\hat{X}_{0})$ or $\hat{V}_{2}(\hat{X}_{0})$, where
$\hat{V}_{1}(\hat{X}_{0})$ is the estimated variance of (\ref{veurachem}),
(\ref{vhomo}) or (\ref{vxoif}) and $\hat{V}_{2}(\hat{X}_{0})$ is the estimated
variance of (\ref{vxoet}). The theoretical variances of $\hat{X}_{0}$ denoted as
$V_{1}(\hat{X}_{0})$ and $V_{2}(\hat{X}_{0})$, are referred, respectively, to the
expressions (\ref{veurachem}), (\ref{vhomo}) or (\ref{vxoif}) and
(\ref{vxoet}) evaluated on the relevant parameter values. In Appendix B it is presented the simulation results.

Tables \ref{xod1p}1, \ref{xod2p}2, \ref{xoc1p}5 and \ref{xoc1g}6 present the
empirical bias, the empirical mean squares error, the theoretical variance and the
estimated variance of $X_{0}$. 
In these tables, it is considered only the variance
(\ref{veurachem}) of the usual model, because based on a simulation study in
\cite{bet} it was shown that the variances (\ref{veurachem}) and (\ref{vshuk}) give
similar results.

Tables \ref{pd1p}3, \ref{pd1g}4 and \ref{pcp}7 present the covering percentages and
the confidence interval amplitudes constructed with a 95\% confidence level for the
parameter $X_{0}$. In Table \ref{pd1p}3, the covering percentages $\%_{1}$ and $\%_{2}$ and amplitudes $A_{1}$ and $A_{2}$ are referred to the confidence intervals constructed using the equations
(\ref{vhomo}) and (\ref{vxoif}).

Tables \ref{xod1p}1-\ref{pd1g}4 consider the homoscedastic
controlled calibration model assuming that $\sigma_{\delta}^{2}$
is unknown.

In Table \ref{xod1p}1 the empirical bias and MSE of $\hat{X}_{0}$ are little and an addition in the size of the variance $\sigma_{\delta}^{2}$,
described in Table \ref{xod2p}2, causes an increasing in the bias and MSE.
Moreover, we have that the bias and MSE of $\hat{X}_{0}$ are smaller when $X_{0}$ is
near to the center value of the variation interval of the variable $X$. These tables
show that for all $n,k$ and $X_{0}$, the theoretical variances obtained using the
expressions (\ref{vxoif}) and (\ref{vxoet}) are equal. This fact occurs also
for the mean estimated variances. We verify also that when $n\geq 20$ and
$k\geq 20$ the theoretical variances and the mean estimated variances from the
proposed model are approximately equal. Observing these tables, we can also notice
that there exists differences between the mean estimated variances of the usual and
proposed models.

Analyzing Tables \ref{pd1p}3 and \ref{pd1g}4, we observe that for all $n$ and
$X_{0}$ when it is adopted erratically the usual model, the amplitudes decrease very
much as the size of $k$ increases. This causes the covering percentage
 to decrease moving away from 95\%. Whereas, adopting the proposed model it is observed
that when $k$ increases the confidence interval amplitude decreases, but the covering percentages increase approaching 95\%. Notice that the covering percentage $\%_{1}$ and $\%_{2}$
and the amplitudes $A_{1}$ and $A_{2}$ are approximately equal, the amplitudes
are very small for $X_{0}=0.8$. In these tables, we observe that when $k=20$ or
100 and when $n$ increases the amplitudes of the intervals decrease and the covering
percentages approaches 95\%. In most cases, the covering percentage obtained
through the proposed model are greater than that for the usual model results and are close to 95\%.

Tables \ref{xoc1p}5 and \ref{pcp}7 describe the results for the controlled
homoscedastic calibration model with $\sigma_{\delta}^{2}$ known. The iterative
method  Quasi-Newton \cite{press} has been used.

In Tables \ref{xoc1p}5 and \ref{xoc1g}6 we have that the empirical bias and SME
decrease as the size of $n$ or $k$ increase and they are small when $X_{0}$ is
near to the central value of the variation interval, $X_{0}=0.8$. When
$\sigma_{\delta}^{2}$ is small (Table \ref{xoc1p}5), for all $n$ and $k$, the
empirical values of MSE from the usual and proposed model are close to the
theoretical variance, but only the mean estimated variance from the proposed model is
close to the theoretical variance. When $\sigma_{\delta}^{2}$ is large (Table
\ref{xoc1g}6), in general, the empirical MSE and the mean estimated variance from the
usual and proposed model are different, but the values supplied by the proposed model
are very close to the theoretical variance.

Analyzing Table \ref{pcp}7, we can make similar comments to the ones we made
 about Tables \ref{pd1p}3 and \ref{pd1g}4.

\section{Aplication}
In this section we test our model, considering both cases $\sigma_{\delta}$ known and unknown, using the data supplied by the chemical laboratory of the "Instituto de Pesquisas Tecnol\'ogicas (IPT)" - Brasil. We also consider the
usual model in order to observe the performance of the proposed model. Our main interest is
 to estimate the unknown concentration value $X_{0}$ of two samples A and B of the
 chemical elements cromo and cadmium.

Tables \ref{cr0} and  \ref{cd0} present the fixed values of concentration of the
standard solutions and the corresponding intensities for the cromo and cadmium element, respectively, which are supplied by the
plasma spectrometry method. This data is referred to as the first stage of the
calibration model.

Tables \ref{cr01} and \ref{cd01} present the intensities corresponding to 3
sample solutions from the sample A and B. This data is referred to as the second stage of
the calibration model.

%
Tables \ref{cr1} and \ref{cd1} describe the estimates of
$\alpha,\,\beta,X_{0},\,V(\hat{X}_{0}),\,\sigma_{\delta}^{2}$ and the
confidence interval amplitude $U(X_{0})$ from the homoscedastic controlled calibration model of
the samples A and B for the chemical elements cromo and cadmium. The values of the variance $\sigma_{\delta}^{2}$ considered as known are obtained from an external study carried out by the IPT, which are $\sigma_{\delta}^{2}=2,5865E-06$ for the cromo element and $\sigma_{\delta}^{2}=0.0017E+02$ for the cadmium element. As seen in Section 2, in order to obtain the estimates of the parameters $\beta$ and $\sigma_{\epsilon}^{2}$ of the proposed model when $\sigma_{\delta}^{2}$ is known, iterative methods are required. In order to solve the system of equations (\ref{ebsd}) and (\ref{eesd}) it was used the
Quasi-Newton iterative method. It is also presented
the estimates from the usual model. The estimates of the variance of $\hat{X}_{0}$
are computed using the relevant expressions (\ref{veurachem}), (\ref{vhomo}) or (\ref{vxoif}). The amplitude $U(X_{0})$ is given by the product of the squared root
of the estimated variance of $\hat{X}_{0}$ and 1.96.

In Tables \ref{cr1} and \ref{cd1} we can observe that the estimates of $\alpha$
and $\beta$ supplied by the usual model is equal to the proposed model when $\sigma_{\delta}^{2}$ is unknown  and they are equal for samples A and B,
this occurs because the expression of the estimators $\hat{\alpha}$ and $\hat{\beta}$
of both models are equal and they only depend on the first stage of the calibration
model. These estimates are slightly different when compared with the estimates
from the proposed model when $\sigma_{\delta}^{2}$ is known. With respect to the
estimate of $X_{0}$, we observe that there is no difference of the estimates
supplied by the usual and the proposed models of the cromo and cadmium element in
both samples A and B, respectively. The estimates of the concentration of the sample
A, of the elements cromo and cadmium, are outside of the variation range of the
standard solution concentrations. We verify that, except to the sample B of the
cadmium element, the estimates of the variance of $\hat{X}_{0}$ and the amplitude
$U(X_{0})$ from the usual model are greater than the estimates supplied by the both
proposed models.

\begin{table}[!h]
\begin{center}
\caption{Concentration $(mg/g)$ and intensity of the standard
solutions of cromo element.} \label{cr0}
\begin{tabular}{|c|c|}\hline
$X_{i}$&Intensity\\\hline
0,05   &6455,900 \\
0,11   &13042,933 \\
0,26   &32621,733 \\
0,79   &97364,500 \\
1,05   &129178,100\\\hline
\end{tabular}
\end{center}
\end{table}
\begin{table}[!h]
\begin{center}
\caption{Intensity of the sample solutions A and
B of cromo element.} \label{cr01}
\begin{tabular}{|c|c|}\hline
\multicolumn{2}{|c|}{Intensity}\\\hline Sample A& Sample B\\\hline
1465,0   &10173,6\\
 1351,0   &10516,9\\
 1495,6 &10352,2\\\hline
 \end{tabular}
\end{center}
\end{table}
\begin{table}[!h]
\begin{center}
\begin{scriptsize}
\caption{Estimates of $\alpha,\,\beta,\,X_{0},$ $V(\hat{X}_{0})$
and the confidence interval amplitude $U(X_{0})$ from the usual and proposed model for the
samples A and B of cromo element.}\label{cr1}
\begin{tabular}{|c|c|c|c|c|c|c|}\hline
&\multicolumn{3}{c|}{Sample A}&\multicolumn{3}{c|}{Sample
B}\\\cline{2-7} Parameters&Usual-M
&\multicolumn{2}{c|}{Proposed-M}&Usual-M
&\multicolumn{2}{c|}{Proposed-M}\\\cline{3-4}\cline{6-7}
 &&unknown $\sigma_{\delta}^{2}$& known $\sigma_{\delta}^{2}$&& unknown $\sigma_{\delta}^{2}$& known $\sigma_{\delta}^{2}$\\\hline
$\alpha$ &123,574 &123,574 &123,889 &123,574 &123,574 &124,021\\
$\beta$  &1,23E+05 &1,23E+05 &1,23E+05 &1,23E+05 &1,23E+05 &1,23E+05\\
$X_{0}$  &0,011   &0,011  &0,011 &0,083  &0,083 &0,083\\
$V(\hat{X}_{0})$  &9,80E-07    &9,15E-07    &1,35E-06 &1,16E-06
&1,13E-06  &1,71E-06\\
$\sigma_{\delta}^{2}$&- &1,60E-06  &-  &-  &5,48E-07 &-\\
 $U(X_{0})$ &2,55E-03  &2,46E-03  &2,99E-03 &2,77E-03  &2,73E-03  &3,36E-03\\\hline
\end{tabular}
\end{scriptsize}
\end{center}
\end{table}

\newpage
\begin{table}[!h]
\begin{center}
\caption{Concentration $(mg/g)$ and intensity of the standard
solutions of cadmium element.} \label{cd0}
\begin{tabular}{|c|c|}\hline
$X_{i}$&Intensity\\\hline
0,05    &4,89733\\
0,10    &9,706\\
0,25    &23,41333\\
0,73    &69,73 \\
1,01    &96,85667\\\hline
\end{tabular}
\end{center}
\end{table}

\begin{table}[!h]
\begin{center}
\caption{Intensity of the sample solutions A and
B of cadmium element.} \label{cd01}
\begin{tabular}{|c|c|}\hline
\multicolumn{2}{|c|}{Intensity}\\\hline Sample A& Sample B\\\hline
0,679  &5,066  \\
0,6837 &5,027  \\
0,6846 &5,085  \\\hline
 \end{tabular}
\end{center}
\end{table}

\begin{table}[!h]
\begin{center}
\begin{scriptsize}
\caption{Estimates of $\alpha,\,\beta,\,X_{0},$ $V(\hat{X}_{0})$ and the confidence interval amplitude $U(X_{0})$ from
the usual and homoscedastic models for the samples A and B of cadmium
element.}\label{cd1}
\begin{tabular}{|c|c|c|c|c|c|c|}\hline
&\multicolumn{3}{c|}{Sample A}&\multicolumn{3}{c|}{Sample
B}\\\cline{2-7} Parameters&Usual-M
&\multicolumn{2}{c|}{Proposed-M}&Usual-M
&\multicolumn{2}{c|}{Proposed-M}\\\cline{3-4}\cline{6-7}
 &&unknown $\sigma_{\delta}^{2}$&known $\sigma_{\delta}^{2}$ &&unknown $\sigma_{\delta}^{2}$& known $\sigma_{\delta}^{2}$\\\hline
$\alpha$ &-0,156  &-0,156  &-0,158 &-0,156  &-0,156  &-0,158\\
$\beta$  &95,828  &95,828  &95,831 &95,828  &95,828  &95,831\\
$X_{0}$  &8,75E-03 &8,75E-03 &8,77E-03 &0,054   &0,054   &0,054\\
$V(\hat{X}_{0})$  &4,06E-06  &3,72E-06  &1,26E-06 &3,81E-06
&3,32E-06    &1,17E-06\\
$\sigma_{\delta}^{2}$&-   &8,31E-06 &-   &-   &8,24E-06 &- \\
 $U(X_{0})$ &5,18E-03  &4,96E-03   &2,89E-03
&5,02E-03   &4,68E-03  &2,78E-03\\\hline
\end{tabular}
\end{scriptsize}
\end{center}
\end{table}

\section{Concluding remarks}
In general, the simulation study reveals that the proposed model is sensible to the presence of error related to the independent variable and gives better results in contrast to the usual  model results. It was noticed that when the error variance $\sigma_{\delta}^{2}$ increases, the mean estimated variance of $\hat{X}_{0}$ obtained using the usual model moves away from the theoretical value. In the example above, the confidence interval amplitude from the proposed models are supplied by the incorporation of error due to the lecture of equipment and the preparation of the standard solutions. It is observed that despite the classical model only considers the error originated from the lecture of the equipment, the amplitude is greater than the obtained by the new approach.\\

\noindent {\bf Acknowledgements}

Betsab\'e G. Blas Achic has been partially supported by IPT (S\~ao
Paulo).
\newpage
\begin{center}\Large\textbf{Appendix}\end{center}
\appendix
\section{Bias and variance for the maximum li\-ke\-lihood\- estimator}
In the following we derive the bias (\ref{Biasxoet}) and the variance (\ref{vxoet}) of the estimator $\hat{X}_{0}$ from the homoscedastic controlled calibration model when $\sigma^{2}_{\delta}$ is known.

Considering the model (\ref{mc1}) and (\ref{mc2}), the estimator $\hat{X}_{0}=(\bar{Y}_{0}-\hat{\alpha})/\hat{\beta}$ can be expressed as
\begin{equation}
\label{xomm}
\hat{X}_{0}=\bar{X}+\frac{\beta(X_{0}-\bar{X})+\bar{\epsilon}_{0}-\bar{\phi}}{\hat{\beta}},
\end{equation}
where $\bar{\epsilon}_{0}=\sum_{i=n+1}^{n+k}\epsilon_{i}/k$ and
$\bar{\phi}=\sum_{i=1}^{n}(\epsilon_{i}-\beta\delta_{i})/n$.

Considering $k$ fixed, expanding $1/\hat{\beta}$ in a Taylor series around $\beta$ and ignoring terms of order less than $n^{-2}$, we obtain the expected value of (\ref{xomm}), given by
\begin{equation}
\nonumber
E(\hat{X}_{0})=X_{0}+\frac{\gamma(\bar{X}-X_{0})}{n\beta^{2}S_{XX}}.
\end{equation}
From this last equation we get the bias (\ref{Biasxoet}).

To derive the variance (\ref{vxoet}) we take the variance of (\ref{xomm}), which is given by
\begin{equation}
\label{Vxomm}
V(\hat{X}_{0})=\beta^{2}(X_{0}-\bar{X})^{2}V(\frac{1}{\hat{\beta}})+V(\frac{\bar{\epsilon}_{0}}{\hat{\beta}})+V(\frac{\bar{\phi}}{\hat{\beta}}).
\end{equation}
We call attention to the fact that (\ref{Vxomm}) is only expressed as a function of the related variances because the corresponding covariances are zero. The variances $V(1/\hat{\beta}), V(\bar{\epsilon}_{0}/\hat{\beta})$ and $V(\bar{\phi}/\hat{\beta})$ can be obtained by expanding $1/\hat{\beta},\,\bar{\epsilon}_{0}/\hat{\beta}$ and $\bar{\phi}/\hat{\beta}$ in a Taylor series around $\beta$ and ignoring terms of order less than $n^{-2}$. They are given by
\begin{eqnarray}
\label{fve1}
V(1/\hat{\beta})&=&\frac{V(\hat{\beta})}{\beta^{4}},\\
\label{fve2}
V(\bar{\epsilon}_{0}/\hat{\beta})&=&\frac{\sigma_{\epsilon}^{2}}{k\beta^{2}}+3\frac{\sigma_{\epsilon}^{2}}{k\beta^{4}}V(\hat{\beta}),\\
\label{fve3}
V(\bar{\phi}/\hat{\beta})&=&\frac{\gamma}{n\beta^{2}}.
\end{eqnarray}
Substituing (\ref{fve1}), (\ref{fve2}) and (\ref{fve3}) in (\ref{Vxomm}), then, the variance (\ref{vxoet}) is obtained.

\newpage
\section{Tables}

Table B1. Empirical bias and mean squared error, theoretical
variance and the mean estimated variance of $\hat{X}_{0}$, for
$\sigma_{\delta}^{2}=0,01$ and unknown.
\begin{center}
\label{xod1p}
\begin{scriptsize}
\begin{tabular}{r|r|r|cc|cc|ccc}\hline
&&&\multicolumn{2}{c|}{Empirical} &
\multicolumn{2}{c|}{Theoretical}&\multicolumn{3}{c}{Mean of
$\hat{V}(\hat{X}_{0})$ }\\ \cline{6-10} $X_{0}$ & $n$ &$k$&&&\multicolumn{2}{c|}{Proposed-M}&\multicolumn{1}{c|}{Usual-M}&\multicolumn{2}{c}{Proposed-M}\\
\cline{4-10} &&& Bias & MSE &$V_{1}(\hat{X}_{0})$ &
$V_{2}(\hat{X}_{0})$&
$\hat{V}_{1}(\hat{X}_{0})$&$\hat{V}_{1}(\hat{X}_{0})$&$\hat{V}_{2}(\hat{X}_{0})$\\\hline
0,01    &5  &2  &-0,0060    &0,0180     &0,0170 &0,0170 &0,0120 &0,0100 &0,0100 \\
    &   &20 &-0,0087    &0,0130     &0,0120 &0,0120 &0,0072 &0,0120 &0,0120 \\
    &   &100    &-0,0060    &0,0130     &0,0120 &0,0120 &0,0065 &0,0120 &0,0120 \\
    &20 &2  &-0,0038    &0,0086     &0,0087 &0,0087 &0,0120 &0,0052 &0,0053 \\
    &   &20 &-0,0028    &0,0043     &0,0042 &0,0042 &0,0033 &0,0040 &0,0040 \\
    &   &100    &-0,0032    &0,0038     &0,0038 &0,0038 &0,0022 &0,0036 &0,0036 \\
    &100    &2  &-0,0023    &0,0058     &0,0058 &0,0058 &0,0100 &0,0027 &0,0027 \\
    &   &20 &-0,0002    &0,0013     &0,0013 &0,0013 &0,0016 &0,0012 &0,0012 \\
    &   &100    &-0,0007    &0,0008     &0,0009 &0,0009 &0,0007 &0,0009 &0,0009 \\\hline
0,8 &5  &2  &-0,0011    &0,0094     &0,0093 &0,0094 &0,0079 &0,0045 &0,0046 \\
    &   &20 &-0,0034    &0,0050     &0,0048 &0,0048 &0,0029 &0,0045 &0,0046 \\
    &   &100    &-0,0007    &0,0047     &0,0044 &0,0044 &0,0024 &0,0045 &0,0045 \\
    &20 &2  &0,0005 &0,0063     &0,0061 &0,0061 &0,0095 &0,0028 &0,0029 \\
    &   &20 &0,0007 &0,0016     &0,0016 &0,0016 &0,0015 &0,0015 &0,0015 \\
    &   &100    &-0,0001    &0,0012     &0,0012 &0,0012 &0,0008 &0,0012 &0,0012 \\
    &100    &2  &0,0005 &0,0050     &0,0052 &0,0052 &0,0099 &0,0021 &0,0021 \\
    &   &20 &-0,0001    &0,0007     &0,0007 &0,0007 &0,0011 &0,0007 &0,0007 \\
    &   &100    &-0,0003    &0,0003     &0,0003 &0,0003 &0,0003 &0,0003 &0,0003 \\\hline
1,9 &5  &2  &0,0041 &0,0160     &0,0150 &0,0160 &0,0120 &0,0093 &0,0094 \\
    &   &20 &0,0026 &0,0110     &0,0110 &0,0110 &0,0065 &0,0110 &0,0110 \\
    &   &100    &0,0076 &0,0110     &0,0110 &0,0110 &0,0058 &0,0110 &0,0110 \\
    &20 &2  &0,0006 &0,0079     &0,0082 &0,0082 &0,0110 &0,0049 &0,0049 \\
    &   &20 &0,0040 &0,0039     &0,0037 &0,0037 &0,0030 &0,0035 &0,0035 \\
    &   &100    &0,0008 &0,0033     &0,0033 &0,0033 &0,0019 &0,0031 &0,0031 \\
    &100    &2  &0,0020 &0,0057     &0,0057 &0,0057 &0,0100 &0,0025 &0,0025 \\
    &   &20 &0,0003 &0,0012     &0,0012 &0,0012 &0,0015 &0,0011 &0,0011 \\
    &   &100    &0,0003 &0,0008     &0,0008 &0,0008 &0,0006 &0,0008 &0,0008 \\
\hline
\end{tabular}
\end{scriptsize}
\end{center}
\newpage

Table B2. Empirical bias and mean squared error, theoretical
variance and the mean estimated variance of $\hat{X}_{0}$, for
$\sigma_{\delta}^{2}=0,1$ and unknown.
\begin{center}
\label{xod2p}
\begin{scriptsize}
\begin{tabular}{r|r|r|cc|cc|ccc}\hline
&&&\multicolumn{2}{c|}{Empirical} &
\multicolumn{2}{c|}{Theoretical}&\multicolumn{3}{c}{Mean of
$\hat{V}(\hat{X}_{0})$ }\\ \cline{6-10} $X_{0}$ & $n$ &$k$&&&\multicolumn{2}{c|}{Proposed-M}&\multicolumn{1}{c|}{Usual-M}&\multicolumn{2}{c}{Proposed-M}\\
\cline{4-10} &&& Bias & MSE &$V_{1}(\hat{X}_{0})$ &
$V_{2}(\hat{X}_{0})$&
$\hat{V}_{1}(\hat{X}_{0})$&$\hat{V}_{1}(\hat{X}_{0})$&$\hat{V}_{2}(\hat{X}_{0})$\\\hline
0,01    &5  &2  &-0,0510    &0,1000     &0,0700 &0,0710 &0,0770 &0,0680 &0,0690 \\
    &   &20 &-0,0500    &0,0950     &0,0660 &0,0660 &0,0220 &0,0660 &0,0660 \\
    &   &100    &-0,0510    &0,0950     &0,0650 &0,0650 &0,0130 &0,0730 &0,0730 \\
    &20 &2  &-0,0180    &0,0280     &0,0250 &0,0250 &0,0670 &0,0240 &0,0240 \\
    &   &20 &-0,0160    &0,0230     &0,0210 &0,0210 &0,0140 &0,0210 &0,0210 \\
    &   &100    &-0,0170    &0,0230     &0,0200 &0,0200 &0,0055 &0,0210 &0,0210 \\
    &100    &2  &-0,0046    &0,0094     &0,0093 &0,0093 &0,0580 &0,0069 &0,0069 \\
    &   &20 &-0,0026    &0,0048     &0,0048 &0,0048 &0,0082 &0,0048 &0,0048 \\
    &   &100    &-0,0033    &0,0043     &0,0044 &0,0044 &0,0029 &0,0044 &0,0044 \\  \hline
0,8 &5  &2  &-0,0084    &0,0370     &0,0290 &0,0290 &0,0460 &0,0250 &0,0250 \\
    &   &20 &-0,0095    &0,0270     &0,0240 &0,0240 &0,0071 &0,0200 &0,0200 \\
    &   &100    &-0,0072    &0,0290     &0,0240 &0,0240 &0,0037 &0,0200 &0,0200 \\
    &20 &2  &-0,0030    &0,0120     &0,0110 &0,0110 &0,0530 &0,0085 &0,0086 \\
    &   &20 &-0,0040    &0,0068     &0,0066 &0,0066 &0,0061 &0,0064 &0,0064 \\
    &   &100    &-0,0031    &0,0063     &0,0062 &0,0062 &0,0017 &0,0060 &0,0060 \\
    &100    &2  &-0,0011    &0,0063     &0,0062 &0,0063 &0,0550 &0,0038 &0,0038 \\
    &   &20 &-0,0015    &0,0017     &0,0017 &0,0017 &0,0057 &0,0017 &0,0017 \\
    &   &100    &-0,0011    &0,0014     &0,0013 &0,0013 &0,0013 &0,0013 &0,0013 \\ \hline
1,9 &5  &2  &0,0430 &0,1090     &0,0630 &0,0630 &0,0830 &0,0750 &0,0750 \\
    &   &20 &0,0450 &0,0860     &0,0580 &0,0580 &0,0210 &0,0650 &0,0650 \\
    &   &100    &0,0410 &0,0860     &0,0580 &0,0580 &0,0110 &0,0600 &0,0600 \\
    &20 &2  &0,0160 &0,0260     &0,0230 &0,0230 &0,0650 &0,0210 &0,0210 \\
    &   &20 &0,0140 &0,0190     &0,0180 &0,0180 &0,0130 &0,0180 &0,0180 \\
    &   &100    &0,0170 &0,0200     &0,0180 &0,0180 &0,0048 &0,0180 &0,0180 \\
    &100    &2  &0,0050 &0,0088     &0,0087 &0,0088 &0,0570 &0,0063 &0,0063 \\
    &   &20 &0,0030 &0,0043     &0,0042 &0,0042 &0,0078 &0,0042 &0,0042 \\
    &   &100    &0,0020 &0,0039     &0,0038 &0,0038 &0,0026 &0,0038 &0,0038 \\
  \hline
\end{tabular}
\end{scriptsize}
\end{center}

\newpage
Table B3. Covering percentage (\%) and amplitude (A) of the
intervals with a 95\% confidence level for the parameter $X_{0}$,
when $\sigma_{\delta}^{2}=0,01$ and unknown. \label{pd1p}
\begin{scriptsize}
\begin{center}
\begin{tabular}{r|r|r|cc|cccc}\hline
$X_{0}$ & $n$&$k$&\multicolumn{2}{c|}{Usual-M}&\multicolumn{4}{c}{Proposed-M}\\
\cline{4-9} &&& \%&$A$&$\%_{1}$&$A_{1}$&$\%_{2}$&$A_{2}$\\
\hline
0,01    &5  &2  &83,04  &0,40   &79,34  &0,36   &79,37  &0,36   \\
    &   &20 &84,95  &0,32   &91,15  &0,41   &91,15  &0,41   \\
    &   &100    &83,72  &0,31   &92,24  &0,42   &92,24  &0,42   \\
    &20 &2  &96,46  &0,42   &83,48  &0,28   &83,52  &0,28   \\
    &   &20 &90,19  &0,22   &92,47  &0,24   &92,47  &0,24   \\
    &   &100    &86,16  &0,18   &92,78  &0,23   &92,78  &0,23   \\
    &100    &2  &98,84  &0,40   &73,16  &0,19   &73,16  &0,19   \\
    &   &20 &96,71  &0,16   &94,14  &0,14   &94,14  &0,14   \\
    &   &100    &91,90  &0,11   &94,68  &0,12   &94,68  &0,12   \\\hline
0,8 &5  &2  &85,43  &0,32   &74,33  &0,24   &74,39  &0,24   \\
    &   &20 &85,16  &0,21   &91,34  &0,26   &91,34  &0,26   \\
    &   &100    &85,04  &0,19   &92,50  &0,25   &92,50  &0,25   \\
    &20 &2  &97,90  &0,38   &73,55  &0,20   &73,55  &0,20   \\
    &   &20 &93,55  &0,15   &93,89  &0,15   &93,89  &0,15   \\
    &   &100    &86,53  &0,11   &93,12  &0,13   &93,12  &0,13   \\
    &100    &2  &99,41  &0,39   &65,05  &0,16   &65,05  &0,16   \\
    &   &20 &98,54  &0,13   &94,05  &0,10   &94,05  &0,10   \\
    &   &100    &94,56  &0,07   &94,86  &0,07   &94,86  &0,07   \\\hline
1,9 &5  &2  &82,47  &0,39   &78,05  &0,35   &78,13  &0,35   \\
    &   &20 &84,50  &0,31   &90,92  &0,39   &90,92  &0,39   \\
    &   &100    &84,75  &0,29   &92,83  &0,39   &92,83  &0,39   \\
    &20 &2  &96,79  &0,41   &83,09  &0,26   &83,11  &0,26   \\
    &   &20 &91,32  &0,21   &93,29  &0,23   &93,31  &0,23   \\
    &   &100    &86,43  &0,17   &93,32  &0,22   &93,32  &0,22   \\
    &100    &2  &98,88  &0,40   &73,06  &0,19   &73,06  &0,19   \\
    &   &20 &97,12  &0,15   &94,31  &0,13   &94,31  &0,13   \\
    &   &100    &92,56  &0,10   &94,74  &0,11   &94,74  &0,11   \\
\hline
\end{tabular}
\end{center}
\end{scriptsize}

\newpage
Table B4. Covering percentage (\%) and amplitude (A) of the
intervals with a 95\% confidence level for the parameter $X_{0}$,
when $\sigma_{\delta}^{2}=0,1$ and unknown. \label{pd1g}
\begin{scriptsize}
\begin{center}
\begin{tabular}{r|r|r|cc|cccc}\hline
$X_{0}$ & $n$&$k$&\multicolumn{2}{c|}{Usual-M}&\multicolumn{4}{c}{Proposed-M}\\
\cline{4-9} &&& \%&$A$&$\%_{1}$&$A_{1}$&$\%_{2}$&$A_{2}$\\
\hline
0,01    &5  &2  &84,89  &0,94   &80,73  &0,85   &80,79  &0,86   \\
    &  &20 &64,10  &0,51   &82,75  &0,86   &82,75  &0,86   \\
    &  &100    &52,09  &0,39   &82,42  &0,86   &82,42  &0,86   \\
    &20 &2  &99,40  &0,99   &91,04  &0,58   &91,10  &0,58   \\
    &  &20 &87,10  &0,45   &92,60  &0,55   &92,62  &0,55   \\
    &  &100    &65,70  &0,28   &92,16  &0,55   &92,18  &0,55   \\
    &100    &2  &100,00 &0,94   &87,60  &0,32   &87,66  &0,32   \\
    &  &20 &98,82  &0,36   &94,88  &0,27   &94,88  &0,27   \\
    &  &100    &89,12  &0,21   &94,84  &0,26   &94,84  &0,26   \\ \hline
0,8 &5  &2  &90,27  &0,73   &80,63  &0,52   &80,75  &0,52   \\
    &  &20 &64,39  &0,31   &82,68  &0,51   &82,74  &0,51   \\
    &  &100    &50,57  &0,23   &84,08  &0,50   &84,08  &0,50   \\
    &20 &2  &99,92  &0,88   &87,30  &0,35   &87,50  &0,35   \\
    &  &20 &91,80  &0,30   &92,78  &0,31   &92,82  &0,31   \\
    &  &100    &67,46  &0,16   &92,38  &0,30   &92,38  &0,30   \\
    &100    &2  &100,00 &0,91   &77,88  &0,22   &77,98  &0,22   \\
    &  &20 &99,94  &0,29   &94,98  &0,16   &95,04  &0,16   \\
    &  &100    &94,60  &0,14   &94,92  &0,14   &94,92  &0,14   \\ \hline
1,9 &5  &2  &85,63  &0,91   &81,42  &0,81   &81,42  &0,81   \\
    &  &20 &61,71  &0,49   &81,89  &0,82   &81,89  &0,82   \\
    &  &100    &50,71  &0,37   &81,24  &0,81   &81,24  &0,81   \\
    &20 &2  &99,54  &0,97   &91,12  &0,54   &91,18  &0,55   \\
    &  &20 &88,10  &0,43   &92,74  &0,52   &92,76  &0,52   \\
    &  &100    &66,44  &0,26   &92,86  &0,51   &92,86  &0,51   \\
    &100    &2  &100,00 &0,93   &86,38  &0,30   &86,38  &0,30   \\
    &  &20 &99,12  &0,35   &94,76  &0,25   &94,76  &0,25   \\
    &  &100    &89,40  &0,20   &95,00  &0,24   &95,00  &0,24   \\
 \hline
\end{tabular}
\end{center}
\end{scriptsize}


\newpage
Table B5. Empirical bias and mean squared error, theoretical
variance and the mean estimated variance of $\hat{X}_{0}$, for
$\sigma_{\delta}^{2}=0,01$ and known.\label{xoc1p}
\begin{scriptsize}
\begin{center}
\begin{tabular}{r|r|r|cc|cc|c|cc}\hline
&&&\multicolumn{4}{c|}{Empirical} &
\multicolumn{1}{c|}{Theoretical}&\multicolumn{2}{c}{Mean of
$\hat{V}(X_{0})$}\\\cline{4-10}
 $X_{0}$ & $n$ &$k$&\multicolumn{2}{c|}{Usual-M} &
\multicolumn{2}{c|}{Proposed-M}& Proposed-M& Usual-M &
Proposed-M\\
\cline{4-8} &&& Bias & MSE & Bias & MSE&$V(\hat{X}_{0})$&&\\\hline
0,01    &5  &2  &-0,0290    &0,0210 &-0,0280    &0,0210 &0,0170 &0,0180 &0,0160 \\
    &   &20 &-0,0290    &0,0140 &-0,0320    &0,0140 &0,0120 &0,0081 &0,0130 \\
    &   &100    &-0,0240    &0,0140 &-0,0270    &0,0140 &0,0120 &0,0070 &0,0130 \\
    &20 &2  &-0,0081    &0,0091 &-0,0064    &0,0090 &0,0086 &0,0130 &0,0076 \\
    &   &20 &-0,0060    &0,0043 &-0,0072    &0,0043 &0,0041 &0,0034 &0,0041 \\
    &   &100    &-0,0038    &0,0038 &-0,0060    &0,0038 &0,0037 &0,0022 &0,0038 \\
    &100    &2  &-0,0011    &0,0056 &-0,0005    &0,0056 &0,0058 &0,0100 &0,0053 \\
    &   &20 &-0,0002    &0,0013 &-0,0001    &0,0013 &0,0013 &0,0016 &0,0012 \\
    &   &100    &-0,0009    &0,0009 &-0,0012    &0,0009 &0,0009 &0,0007 &0,0009 \\  \hline
0,8 &5  &2  &-0,0074    &0,0110 &-0,0072    &0,0100 &0,0093 &0,0120 &0,0085 \\
    &   &20 &-0,0046    &0,0051 &-0,0051    &0,0052 &0,0048 &0,0032 &0,0049 \\
    &   &100    &-0,0076    &0,0048 &-0,0082    &0,0048 &0,0044 &0,0025 &0,0046 \\
    &20 &2  &-0,0034    &0,0063 &-0,0031    &0,0063 &0,0061 &0,0100 &0,0052 \\
    &   &20 &0,0001 &0,0016 &0,0000 &0,0016 &0,0016 &0,0015 &0,0016 \\
    &   &100    &-0,0009    &0,0012 &-0,0013    &0,0012 &0,0012 &0,0008 &0,0012 \\
    &100    &2  &0,0000 &0,0053 &0,0002 &0,0053 &0,0052 &0,0099 &0,0048 \\
    &   &20 &0,0000 &0,0007 &0,0000 &0,0007 &0,0007 &0,0011 &0,0007 \\
    &   &100    &-0,0001    &0,0003 &-0,0002    &0,0003 &0,0003 &0,0003 &0,0003 \\  \hline
1,9 &5  &2  &0,0200 &0,0180 &0,0200 &0,0180 &0,0150 &0,0170 &0,0140 \\
    &   &20 &0,0240 &0,0120 &0,0260 &0,0130 &0,0110 &0,0071 &0,0120 \\
    &   &100    &0,0200 &0,0130 &0,0230 &0,0130 &0,0100 &0,0062 &0,0110 \\
    &20 &2  &0,0037 &0,0082 &0,0021 &0,0081 &0,0082 &0,0120 &0,0071 \\
    &   &20 &0,0059 &0,0037 &0,0066 &0,0037 &0,0037 &0,0030 &0,0036 \\
    &   &100    &0,0033 &0,0032 &0,0051 &0,0032 &0,0032 &0,0020 &0,0033 \\
    &100    &2  &0,0020 &0,0058 &0,0015 &0,0057 &0,0057 &0,0100 &0,0053 \\
    &   &20 &0,0003 &0,0012 &0,0002 &0,0012 &0,0012 &0,0015 &0,0011 \\
    &   &100    &0,0003 &0,0008 &0,0006 &0,0008 &0,0008 &0,0006 &0,0008 \\
  \hline
\end{tabular}
\end{center}
\end{scriptsize}

\newpage
Table B6. Empirical bias and mean squared error, theoretical
variance and the mean estimated variance of $\hat{X}_{0}$, for
$\sigma_{\delta}^{2}=0,1$ and known.\label{xoc1g}
\begin{scriptsize}
\begin{center}
\begin{tabular}{r|r|r|cc|cc|c|cc}\hline
&&&\multicolumn{4}{c|}{Empirical} &
\multicolumn{1}{c|}{Theoretical}&\multicolumn{2}{c}{Mean of
$\hat{V}(X_{0})$}\\\cline{4-10}
 $X_{0}$ & $n$ &$k$&\multicolumn{2}{c|}{Usual-M} &
\multicolumn{2}{c|}{Proposed-M}& Proposed-M& Usual-M &
Proposed-M\\
\cline{4-8} &&& Bias & MSE & Bias & MSE&$V(\hat{X}_{0})$&&\\\hline
0,01    &5  &2  &-0,4330    &0,5590 &-0,3830    &0,4580 &0,0590 &0,3650 &0,2310 \\
    &   &20 &-0,1140    &0,1310 &0,4140 &1,0880 &0,0540 &0,0250 &0,0760 \\
    &   &100    &-0,1890    &0,1540 &-0,0290    &0,7730 &0,0540 &0,0200 &0,1250 \\
    &20 &2  &-0,0930    &0,0430 &-0,0770    &0,0360 &0,0210 &0,0950 &0,0380 \\
    &   &20 &-0,0490    &0,0210 &-0,0510    &0,0210 &0,0160 &0,0150 &0,0180 \\
    &   &100    &-0,0430    &0,0190 &-0,0250    &0,0640 &0,0150 &0,0058 &0,0170 \\
    &100    &2  &-0,0200    &0,0110 &-0,0160    &0,0097 &0,0084 &0,0630 &0,0110 \\
    &   &20 &-0,0077    &0,0041 &-0,0085    &0,0039 &0,0037 &0,0084 &0,0038 \\
    &   &100    &-0,0066    &0,0037 &-0,0087    &0,0037 &0,0033 &0,0029 &0,0033 \\  \hline
0,8 &5  &2  &-0,0500    &0,0620 &-0,0430    &0,0550 &0,0280 &0,1150 &0,0620 \\
    &   &20 &-0,0450    &0,0380 &0,0820 &0,0810 &0,0240 &0,0086 &0,0320 \\
    &   &100    &-0,0530    &0,0510 &-0,0069    &0,0980 &0,0230 &0,0055 &0,0360 \\
    &20 &2  &-0,0280    &0,0150 &-0,0240    &0,0140 &0,0110 &0,0740 &0,0210 \\
    &   &20 &-0,0079    &0,0068 &-0,0085    &0,0069 &0,0064 &0,0065 &0,0067 \\
    &   &100    &-0,0055    &0,0065 &-0,0030    &0,0079 &0,0060 &0,0018 &0,0062 \\
    &100    &2  &-0,0030    &0,0067 &-0,0021    &0,0065 &0,0062 &0,0600 &0,0091 \\
    &   &20 &-0,0018    &0,0018 &-0,0018    &0,0017 &0,0017 &0,0058 &0,0017 \\
    &   &100    &-0,0021    &0,0013 &-0,0025    &0,0013 &0,0013 &0,0013 &0,0013 \\  \hline
1,9 &5  &2  &0,3480 &0,3770 &0,3070 &0,3000 &0,0530 &0,2850 &0,1760 \\
    &   &20 &0,1400 &0,1630 &-0,3310    &0,9370 &0,0490 &0,0300 &0,0780 \\
    &   &100    &0,0410 &0,0180 &0,0270 &0,0510 &0,0140 &0,0051 &0,0150 \\
    &20 &2  &0,0970 &0,0430 &0,0790 &0,0350 &0,0190 &0,0930 &0,0350 \\
    &   &20 &0,1400 &0,1630 &-0,3310    &0,9370 &0,0490 &0,0300 &0,0780 \\
    &   &100    &0,1670 &0,1300 &-0,0005    &0,7090 &0,0480 &0,0160 &0,1120 \\
    &100    &2  &0,0200 &0,0093 &0,0160 &0,0087 &0,0080 &0,0620 &0,0110 \\
    &   &20 &0,0120 &0,0039 &0,0130 &0,0037 &0,0033 &0,0080 &0,0035 \\
    &   &100    &0,0072 &0,0031 &0,0089 &0,0031 &0,0029 &0,0027 &0,0030 \\
  \hline
\end{tabular}
\end{center}
\end{scriptsize}

\newpage
Table B7. Covering percentage (\%) and amplitude (A) of the
intervals with a 95\% confidence level for the parameter $X_{0}$,
when $\sigma_{\delta}^{2}=0,01.$ and 0,1 and known.\label{pcp}
\begin{scriptsize}
\begin{center}
\begin{tabular}{r|r|r|cccc|cccc}\hline
&&&\multicolumn{4}{c|}{$\sigma_{\delta}^{2}=0,01$}&\multicolumn{4}{c}{$\sigma_{\delta}^{2}=0,1$}\\
\cline{4-11} $X_{0}$ & $n$
&$k$&\multicolumn{2}{c|}{Usual-M}&\multicolumn{2}{c|}{Proposed-M}&\multicolumn{2}{c|}{Usual-M}&\multicolumn{2}{c}{Proposed-M}\\
\cline{4-11}
&&& \% & A  & \%& A&\%& A & \%& A \\
\cline{4-11} \hline
0,01    &5  &2  &92,10  &0,51   &91,20  &0,48   &95,06  &1,89   &92,40  &1,49   \\
    &   &20 &87,32  &0,34   &95,18  &0,44   &63,38  &0,59   &64,47  &1,09 \\
    &   &100    &84,50  &0,32   &95,19  &0,44 &52,69  &0,48   &89,24  &1,18    \\
    &20 &2  &97,46  &0,43   &90,12  &0,33   &99,87  &1,17   &96,27  &0,71 \\
    &   &20 &91,00  &0,23   &94,03  &0,25  &92,76  &0,48   &94,48  &0,52   \\
    &   &100    &86,73  &0,19   &95,21  &0,24   &70,38  &0,30   &90,35  &0,51  \\
    &100    &2  &97,01  &0,43   &90,00  &0,33 &100,00 &0,98   &93,41  &0,40   \\
    &   &20 &92,34  &0,23   &95,26  &0,25  &99,77  &0,36   &95,09  &0,24    \\
    &   &100    &85,93  &0,19   &94,55  &0,24    &92,70  &0,21   &94,42  &0,23   \\ \hline
0,8 &5  &2  &94,33  &0,41   &89,28  &0,35 &98,77  &1,31   &97,54  &0,98   \\
    &   &20 &86,84  &0,22   &95,03  &0,27   &61,85  &0,34   &81,25  &0,69  \\
    &   &100    &85,65  &0,19   &95,56  &0,27    &49,29  &0,27   &87,82  &0,69 \\
    &20 &2  &98,35  &0,39   &88,44  &0,27   &100,00 &1,03   &94,09  &0,53    \\
    &   &20 &93,84  &0,15   &94,41  &0,15   &94,44  &0,32   &95,11  &0,32     \\
    &   &100    &85,98  &0,11   &94,32  &0,14   &70,38  &0,17   &92,41  &0,31   \\
    &100    &2  &98,08  &0,39   &88,22  &0,27   &100,00 &0,95   &92,25  &0,34     \\
    &   &20 &92,75  &0,15   &93,96  &0,15 &99,89  &0,30   &94,89  &0,16\\
    &   &100    &87,17  &0,11   &94,60  &0,14 &95,00  &0,14   &94,88  &0,14 \\ \hline
1,9 &5  &2  &92,30  &0,50   &90,61  &0,46   &96,26  &1,78 &92,14
&1,45\\
    &   &20 &86,96  &0,33   &95,13  &0,42  &63,58  &0,56   &67,24  &1,03   \\
    &   &100    &85,84  &0,30   &94,73  &0,42   &49,30  &0,46   &83,75  &1,21   \\
    &20 &2  &97,61  &0,43   &89,92  &0,32  &100,00 &1,17   &96,27  &0,70   \\
    &   &20 &91,60  &0,21   &94,27  &0,23   &94,26  &0,46   &94,95  &0,49  \\
    &   &100    &86,16  &0,17   &94,99  &0,23   &72,44  &0,28   &94,18  &0,48   \\
    &100    &2  &97,04  &0,43   &89,70  &0,32   &100,00 &0,97   &94,27  &0,39  \\
    &   &20 &91,21  &0,21   &93,85  &0,23   &99,59  &0,35   &95,18  &0,23   \\
    &   &100    &87,04  &0,17   &95,11  &0,23   &93,38  &0,20   &94,98  &0,21   \\
 \hline
\end{tabular}
\end{center}
\end{scriptsize}

\end{document}